\documentclass[prd,amsmath,notitlepage,twocolumn]{revtex4-1}
\usepackage{graphicx}
\usepackage{float}
\usepackage{epstopdf,cancel}
\usepackage{epsf,latexsym,bbm,euscript}
\usepackage{amssymb,amsmath}
\usepackage{mathtools} 
\usepackage{times,graphics}
\usepackage{soul,xcolor}

\def\beq{\begin{equation}}
\def\eeq{\end{equation}}

\def\be{\begin{equation}}
\def\ee{\end{equation}}

\def\6{\langle}
\def\9{\rangle}

\def\bali{\begin{align}}
\def\eni{{\end{align}}}
   \def\sg{\textsl{g}}

\begin{document}

\title{Trapped surfaces, energy conditions, and horizon avoidance in spherically-symmetric collapse}
\author{Valentina Baccetti }
\address{School of Science, RMIT University,\\
Melbourne, VIC 3000, Australia}

\author{Robert B. Mann}

\address{Department of Physics and Astronomy, University of Waterloo,\\
Waterloo, ON,  N2L 3G1, Canada}
\address{Perimeter Institute for Theoretical Physics,\\ Waterloo, ON, N2L 6B9, Canada}

\author{Daniel R. Terno$^*$}

\address{Department of Physics and Astronomy, Macquarie University,\\
Sydney, NSW 2109, Australia\\
E-mail: daniel.terno@mq.edu.au}

\begin{abstract}

We consider  spherically-symmetric black holes in semiclassical gravity. For   a collapsing   radiating thin shell we derive a sufficient condition on the exterior geometry
 that ensures that a black hole is not formed. This is also a sufficient condition for an infalling  test particle to
  avoid the apparent horizon of an existing black hole and approach it only within a certain minimal distance.
  Taking the presence of a trapped region and its outer  boundary --- the apparent horizon--- as the defining feature of black holes, we explore the consequences of their
  finite time of formation according to a distant observer.
Assuming regularity of the apparent horizon we obtain the limiting form of the metric and the energy-momentum tensor in its vicinity that violates the null energy condition (NEC).
 The metric does not satisfy the sufficient condition for  horizon avoidance: a thin  shell collapses to form a black hole and test particles (unless too slow)  cross into it in finite time. However, there may be difficulty in maintaining the expected range of the NEC violation, and stability against perturbations is not assured.

\vspace{5mm}

\end{abstract}

\vspace{5mm}

\maketitle 
\section{Introduction}

Event horizon --- the null surface that bounds the spacetime region from which signals cannot escape  --- is the defining feature of black holes in general relativity
\cite{he:book,poisson,bambi}. This classical concept plays an important role in their quantum {behaviour} \cite{bmps:95,kiefer:07,modern}.
Emission of the Hawking radiation completes a thermodynamic picture of black holes, and its most straightforward derivation relies on the existence of an horizon \cite{bmps:95}.
This radiation is also one of the ingredients of the black hole information loss paradox \cite{h:76}, perhaps  the longest-running controversy in theoretical physics  \cite{kiefer:07,modern,h:76}.

 Event horizons are global teleological entities that are in principle unobservable
   \cite{visser:14,cp:na17}.  Theoretical, numerical and observational studies therefore focus on other characteristic features of black holes \cite{bambi,faraoni:b}.
A {local} expression of the idea of absence of communications with the outside world is provided  by the notion of a trapped region. It is a domain
where both outgoing and ingoing future-directed null
  geodesics emanating from a spacelike  two-dimensional surface  with spherical topology have negative expansion \cite{he:book,bambi,faraoni:b,krishnam:14}. The apparent horizon is the outer
boundary of the trapped region \cite{he:book, faraoni:b}.  According to classical general relativity the  apparent horizon is located inside the event
horizons if the matter  satisfies energy conditions \cite{he:book,econ}.

Quantum states can violate energy conditions \cite{econ}. 
Black  hole evaporation proceeds precisely because $T_{\mu\nu}=\6\hat T_{\mu\nu}\9$ violates the null energy condition (NEC):  there is a null vector $k^\mu$ such that $T_{\mu\nu}k^\mu k^\nu<0$.
In this case the apparent horizon is outside the event horizon. In fact, the very existence of the latter is uncertain \cite{ab:05,nohorsin}.
 While existence of
 spacetime singularities is no longer prescribed,
their appearance without the horizon cover (``naked'') is not excluded either.
This situation motivated introduction of many models of the ultra-compact objects \cite{cp:na17}.

According to a distant observer   formation of a classical black holes takes an infinite amount of time $t$, even if effective
blackening out happens very fast. Similarly, the plunge of a test particle into an existing black holes takes infinite amount of time $t$, but finite proper time of a comoving observer. On the
other hand, if quantum effects responsible for finite-time black hole evaporation allow for the formation of an apparent horizon, it happens in
finite $t$. The question is then if it is possible to fall into such a black hole.

Working in the framework of semiclassical gravity with
spherical symmetry \cite{bmt:18}, we consider formation of a trapped region in the finite time of a distant observer as a  definition of existence of a black hole.
Assuming that it exists, we derive the condition that allows an infalling observer to avoid horizon crossing. We then show that if the apparent horizon is regular, the energy-momentum tensor and
the metric in its neighbourhood are uniquely defined by the Schwarzschild radius $r_\sg$ and its rate of change. The resulting metric does not satisfy the sufficient condition for  horizon avoidance.
 Finally, we discuss intriguing implications of these results.
\section{Spherical symmetry. Sufficient condition for horizon avoidance}

A general spherically symmetric metric in Schwarzschild coordinates is given by
\be
ds^2=-e^{2h(t,r)}f(t,r)dt^2+f(t,r)^{-1}dr^2+r^2d\Omega \label{sgenm}
\ee
The function $f(t,r)=1-C(t,r)/r$ is coordinate-independent, where the function $C(t,r)$ is the Misner-Sharp mass \cite{bambi,aphor}. In an asymptotically flat
spacetime $t$ is the physical time of a distant observer.

Trapped regions exist only if the equation $f(t,r)=0$ has a root \cite{krishnam:14}.
This root (or, if there are several, the largest one) is the  {Schwarzschild   radius} $r_\sg(t)$.  Apparent horizons are in general observer-dependent  entities. However they are unambiguously defined in the spherically symmetric case for all spherical-symmetry preserving foliations\cite{aphor}. In this case the apparent horizon is located at $r_\sg$.  In the Schwarzschild spacetime $C(t,r)=2M$ and $h= 0$, hence $r_\sg=2M$.

In  thin shell collapse models \cite{poisson} the geometry inside the shell is given by the flat Minkowski metric. The matter content of the shell
 is given by the surface energy-momentum tensor. The trajectory of a massive shell is parameterized by its proper time $\tau$ and expresses as $\big(T(\tau), R(\tau)\big)$ in
 the exterior Schwarzschild coordinates. Initially the shell is located outside its gravitational radius, $R(0)>r_\sg$.
Its dynamics is obtained by using the so-called junction conditions \cite{poisson, isr:66}.

The first junction condition  is the statement that the induced metric $h_{ab}$ on the shell $\Sigma$ is the same on  both sides $\Sigma^\pm$,
$ds^2_\Sigma=h_{ab}dy^ady^b=-d\tau^2+R^2d\Omega$. Since for massive particles the four-velocity $u^\mu$ satisfies $u_\mu u^\mu=-1$, the shell's trajectory obeys
\be
\dot T=\frac{\sqrt{F+\dot R^2}}{e^HF}, \label{udot}
\ee
where   $\dot A=dA/d\tau$, $H=h(T,R)$, $F=f(T,{R})$.   This condition is used to identify the radial coordinate of the shell in interior and exterior coordinates, $R_-\equiv R$.

Discontinuity of the extrinsic curvature $K_{ab}$ is described by the second junction condition \cite{poisson,isr:66} that relates it to the surface
 energy-momentum tensor. Given the exterior metric the junction conditions result in the equations of motion for the shell.    For a classical collapse in  vacuum   the exterior geometry
 is given by the Schwarzschild metric, and the resulting equation
for $R(\tau)$ is simple enough  to have an analytic solution $\tau(R)$, leading to the finite proper time $\tau(r_\sg)$ and infinite time $T(r_\sg)$.

This equation of motion is modified for a general exterior metric and its solution has some remarkable features \cite{bacu:18,mnt:18}.
Here we focus on the possibility of crossing the Schwarzschild sphere of an evaporating black hole ($r_\sg'(t)<0$  )
in finite proper time.
For a finite evaporation time $t_E$ the finite proper crossing time is equivalent to having  a finite  time $t_\sg$ of a distant observer.
  By monitoring the gap between the shell and the Schwarzschild radius \cite{bmt:18,kmy:13},
\be
X(\tau):=R(\tau)-r_\sg\big(T(\tau)\big),
\ee
we discover  the sufficient condition for a thin shell to never cross its Schwarzschild radius. The same condition applies to the study of
{an infalling test particle} into an existing black hole. The analysis
is  generalized  to null shells and test particles.

The rate of approach to the Schwarzschild radius  behaves as
\be
\dot X= \dot R  -r'_\sg(T)\dot T. \label{xeps}
\ee
Close to the Schwarzschild radius we have $\dot T\approx-\dot R e^{-H}/F$, and hence
\be
\dot X\approx \dot R(1-|r_\sg'|e^{-H}/F).
\ee
If for a fixed $t$ the function $\exp(h)f$   goes to zero as $x:=r-r_\sg\to 0$, then there is a stopping scale $\epsilon_*(\tau)$. If the shell comes to the Schwarzschild radius closer than $\epsilon_*$
the gap  has to increase, $\dot X>0$, {evidently indicating in this case that}
 the shell never collapses  to a black hole.  It is so if, e.g., $h(t,r)\leq 0$.

 In particular,
  this is the case when the exterior geometry is given by the outgoing Vaidya metric. Then $\epsilon_*=2C|dC/dU|$, where $U(\tau)$ is the retarded null coordinate of the shell \cite{bmt:18,kmy:13}.
 However,
it is not a priori clear that in a general evaporating case  this criterion is satisfied.

\section{Metric outside an apparent horizon}

Using only one additional assumption it is possible to obtain the explicit form of the metric near $r_\sg$. In fact  this metric satisfies the sufficient condition for the horizon avoidance.
 We consider an evaporating black hole that is formed {at some distant observer's finite time}, i.e. its apparent horizon radius $r_\sg(t)$ is a decreasing function of time.
In addition we assume that the horizon is regular (the standard ``no drama at the horizon'' postulate\cite{modern}, where the established regularity of the classical results is assumed to hold in the quantum-dominated regime).
The regularity is expressed by finite values of the curvature scalars  that can be directly expressed in terms of the energy-momentum tensor\cite{bmmt:18}, $\mathrm{T}:=T^{\mu}{}_{\mu}$ and $\mathfrak{T}:=T^{\mu\nu}T_{\mu\nu}$.

The existence  {of an apparent horizon} and regularity assumptions strongly constrain the energy-momentum tensor, and consistency with the known results on the background of an eternal black hole\cite{bmps:95,leviori:16}
specify its limiting form uniquely. The leading terms in the $(tr)$ block of the energy-momentum tensor turns out to be the functions of $a^2:=r_\sg'^2 r_\sg$, and  {the functions $C(t)$ and $h(t)$ take the following form} \cite{bmmt:18},
 \be
C= r_\sg(t)-a(t)\sqrt{x}+\frac{1}{3}x\ldots.  \label{c0sin}
\ee
and
 \be
h=-\ln\frac{\sqrt{x}}{\xi_0(t)} +\frac{4}{3a}\sqrt{x}+\ldots,\label{ho}
\ee
where $x=r-r_\sg$.   The function of time {$\xi_0(t)$} is   determined by the  choice of the time variable. In asymptotically flat spacetimes with $t$ being a physical
time of a stationary Bob,  {$\lim_{r\to \infty} h(t,r)=0$}
and the exact solution of the Einstein equations allows to determine  {$\xi_0$.} The constant in the function $h$ is set (using the freedom in re-defining the time variable)
in such a way that $h\approx 0$ for a macroscopic BH when $x\to r_\sg$,
  i.e. far relative to the scale of quantum effects $\alpha^2$.   The metric takes a particularly simple form in  ingoing Vaidya coordinates\cite{bmmt:18}.

  The energy-momentum tensor that corresponds to this metric violates the null energy condition in the vicinity of the  apparent horizon\cite{bmmt:18}.
  The comoving density and pressure at the apparent horizon of an evaporating black hole are negative,
\be
\rho=p=-\frac{r_\sg'^2}{16\pi \dot r^2 r_\sg^2},
\ee
 where $r$ is the radial coordinate of the comoving observer\cite{t:19}.
  We focus on the question of horizon avoidance. Return first to a collapsing thin shell problem where the exterior metric is given now by Eq.~\eqref{sgenm} with the metric functions that are given above.
  Expanding  Eqs.~\eqref{c0sin} and \eqref{ho}  for $X\to 0$  the rate of approach
    \be
    \dot X=-\frac{(\dot R^2-4\pi r_\sg^2\Upsilon^2)}{2|\dot R|\sqrt{\pi}r_\sg^{3/2}\Upsilon}\sqrt{X}+\ldots,
    \ee
which is independent of the function $\xi_0$. Hence if a test particle is in the vicinity of the apparent horizon, $X\ll a^2$,
 it will cross the horizon unless $|\dot R| <2\sqrt{\pi}r_\sg\Upsilon\sim \sqrt{\kappa}\sim 0.01$.

For the evaporating case we match the $r_\sg'$ that is obtained as a consistency requirement on the functions of $C$ and $h$ from the Enstein equations \cite{bmmt:18},
with the known results~\cite{bmt:s18} for the quasi-static mass loss
$r_\sg'=-\kappa/r_\sg^2$. In Planck units $\kappa\sim 10^{-3} - 10^{-4}$, and we obtain \be
\xi_0 \approx \sqrt{\frac{\kappa}{2 r_\sg}} \approx2\sqrt{\pi r_\sg^3}\Upsilon = \frac{a}{2}.
\ee

We have seen that the violation of the NEC is necessary for existence of a black hole. Such violations, however, are bounded by quantum energy inequalities (QEI). Adapting the QEI of Ref.~\citenum{fp:06}
we obtain that the region where the NEC is violated is bounded by \be
x_{\max}<\frac{11}{960 \pi |r_\sg'|r_\sg^2}\sim 1, \ee
that is obtained by ignoring the sub-Planckian features of the bound\cite{bmt:18}. Even this estimate fails short of the conventional estimate  $x_{\max}\sim r_\sg$.
Our results indicate either that the required negative energy density for
having a  Schwarzschild sphere at finite time $t_\mathrm{S}$ cannot be maintained or the trapped regions forms at an much later stage of the collapse. Alternatively, the domain of validity of our metric (that
has the form\cite{bmt:18} of the approximate metric that is obtained by taking the backreaction into account\cite{bmps:95} has a much narrower domain of validity.
 Both possibilities may indicate that the semiclassical approximation and its associated classical notions are modified already at the horizon or larger scales. A rigourous analysis of this situation is in order.

 Another question results from two properties we discussed above. On the one hand, existence of the apparent horizon requires violation of the NEC. Test particles cross it in finite time unless moving too slow.
 It is not clear how the apparent horizon that requires  NEC violation for its existence reacts to a perturbation by infalling normal matter. Given that collapsing thin shells are known to cross
 the apparent horizon and form a black hole with
 nearly all their rest mass intact\cite{bmt:s18}, it is possible that the answer will involve considerable adjustment of our concept of black holes.

\section*{Acknowledgements}
The work of RBM was supported in part by the Natural Sciences and Engineering Research Council of Canada and the Distinguished Visitor Program of the Faculty of Science and Engineering of the Macquarie University.
 We thank Sabine Hossenfelder, Jorma Luoko, Pady Padmanabhan, Amos Ori, Don Page,  Bill Unruh, Matt Visser, and Mark Wardle for useful discussions and critical comments.

\end{document}